\documentclass[11pt]{article}
\usepackage[dvips]{graphicx}
\usepackage{amssymb}
\usepackage{amsmath}
\usepackage{amscd}
\usepackage{mathrsfs}
\usepackage{latexsym}
\usepackage{makeidx}

\date{}
\author{Claudio Garola\footnote{E-mail: garola@le.infn.it} \ \ and \ 
Sandro Sozzo\footnote{E-mail: sozzo@le.infn.it} \\ Dipartimento di Fisica and Sezione INFN \\ Universit\`a del Salento, via Arnesano, 73100 Lecce, Italy}
\title{\textbf{Embedding Quantum Mechanics Into a Broader Noncontextual Theory: A Conciliatory Result}}

\begin{document}
\maketitle

\begin{abstract}
\noindent
The \emph{extended semantic realism} (\emph{ESR}) \emph{model} embodies the mathematical formalism of standard (Hilbert space) quantum mechanics in a noncontextual framework, reinterpreting quantum probabilities as \emph{conditional} instead of \emph{absolute}. We provide here an improved version of this model and show that it predicts that, whenever idealized measurements are performed, a \emph{modified Bell--Clauser--Horne--Shimony--Holt} (\emph{BCHSH}) \emph{inequality} holds if one takes into account all individual systems that are prepared, standard quantum predictions hold if one considers only the individual systems that are detected, and a \emph{standard BCHSH inequality} holds at a microscopic (purely theoretical) level. These results admit an intuitive explanation in terms of an unconventional kind of unfair sampling and constitute a first example of the unified perspective that can be attained by adopting the ESR model.

\vspace{.3cm}

\vspace{.3cm}
\end{abstract}

\section{Introduction\label{intro}}
It is well known that the \emph{objectification problem} is crucial in the quantum theory of measurement. According to Busch, Lahti and Mittelstaedt \cite{blm91} this problem arises whenever an interpretation of the mathematical apparatus of standard (Hilbert space) quantum mechanics (QM) is adopted which is \emph{realistic} (in the sense that it assumes that QM deals with individual objects and their properties) and \emph{complete} (in the sense that it assumes that all elements of physical reality are described by QM).  

If one wants to trace the origins of the objectification problem it is expedient to introduce the notion of \emph{objectivity}. To be precise, we say that in a physical theory $\mathcal T$ which is realistic in the sense explained above a property $E$ of a physical system $\Omega$ is objective for a given state $S$ of $\Omega$ if, for every individual example $x$ of $\Omega$ (\emph{physical object}) in the state $S$, $E$ is either possessed or not possessed by $x$, independently of any measurement that may be performed on $x$;\footnote{The terms ``possessed'' and ``not possessed'' in the definition above are rather loose. We show in the following that our definition of objectivity acquires a more precise meaning whenever its implications are considered on the semantics of the observational language of QM or on the properties of hidden--variables theories for QM.} furthermore, we say that $\mathcal T$ is an objective theory if all physical properties are objective in it for every state $S$ of $\Omega$, nonobjective otherwise. Then, there exist some celebrated ``no--hidden--variables'' theorems (in particular, the Bell--Kochen--Specker, or Bell--KS, and the Bell theorems \cite{b66,ks67,b64}) which according to most scholars show that QM is a \emph{contextual}, hence nonobjective, theory. In particular, if a pure state $S$ is given, a physical property $E$ is nonobjective in $S$ if and only if one has probability different from 1 or 0 of finding the property $E$ when performing a measurement on a physical object $x$ in the state $S$. If one then maintains that QM is a complete theory, one cannot explain how physical properties that are not objective may become objective (and conversely) when a measurement is performed, which is just the objectification problem.

According to the authors quoted above, no satisfactory solution of the objectification problem has been found in the framework of the realistic and complete interpretation of QM or its unsharp extension \cite{blm91,bs96}. Therefore, many proposals have been made to avoid it, some of which consider QM as an incomplete theory. In this case one could contrive a hidden--variables theory containing parameters whose unknown values determine whether a property is possessed or not by a given physical object $x$ in the state $S$, hence determine the outcome of any measurement on $x$. The objectification problem thus disappears. But the ``no--hidden--variables'' theorems oblige one to introduce hidden variables whose values depend on the measurement context and not only on the features of the physical object that is considered (\emph{e.g.}, Bohm's theory), hence objectivity of physical properties is lost anyway. Moreover, Bell's theorem implies that contextuality holds also at a distance (\emph{nonlocality}) \cite{b64,m93}, which is strongly counterintuitive because it implies that performing a measurement on a part of a physical system may make objective a property of another far away part of the physical system which was previously nonobjective. In addition, from a semantic point of view nonobjectivity implies that a statement $E(x)$ attributing a physical property $E$ to a physical object $x$ in a state $S$ has a truth value if and only if $E$ is objective in the state $S$: this amounts to adopt a verificationist theory of truth for the observational language of QM, which is highly problematical \cite{gs04}.

Because of the consequences of nonobjectivity mentioned above one may wonder whether an interpretation of the mathematical apparatus of QM can be provided which recovers objectivity of physical properties. But, of course, this possibility seems to be excluded by the ``no--hidden--variables'' theorems. Hence every attempt at vindicating objectivity of QM must begin with a preliminary criticism of these theorems. Bearing in mind this remark, one of the authors, together with some coworkers, has proven in several papers (see, \emph{e.g.}, \cite{ga00,gs96a,gs96b,ga02b}) that the standard reasonings aiming to show that QM is a contextual theory accept implicitly an epistemological assumption (called \emph{metatheoretical classical principle}, or, briefly, MCP) which does not fit in well with the operational philosophy of QM. If MCP is replaced by a weaker \emph{metatheoretical generalized principle} (MGP), which is closer to the aforesaid philosophy, the proof of the conflict of QM with objectivity cannot be given.\footnote{For the sake of completeness let us discuss this issue in more details. To this end, let us resume the essentials of our epistemological position. We consider the \emph{theoretical laws} of QM as mathematical schemes from which \emph{empirical laws} can be deduced. Consistently with the operational and antimetaphysical attitude of QM, we do not attribute truth values to the sentences stating the former laws. We instead assume that every sentence stating an empirical law has a truth value, which is \emph{true} in all those situations in which the law can be checked (\emph{epistemically accessible physical situations}), while it may be \emph{true} as well as \emph{false} in physical situations in which it cannot be checked because QM itself prohibits any test. This assumption constitutes the general principle that we call MGP. If one then considers the ``no--hidden--variables'' theorems mentioned above, one sees that they are proved \emph{ab absurdo}. To be precise, one assumes boundary, or initial, conditions which attribute noncompatible properties to the physical system that is considered. This implies hypothesizing physical situations that are not epistemically accessible. Nevertheless empirical quantum laws are applied in these situations \cite{ga02,gp04}, which implies assuming that empirical laws are valid in QM independently of the epistemic accessibility of the physical situation that is considered. This assumption, which is stronger than MGP, constitutes the general principle that we call MCP. If MCP is rejected and replaced by MGP, the ``no--hidden--variables'' theorems cannot be proved. \label{mgp}} Basing on this result and adopting MGP in place of MCP, a new interpretation of the mathematical formalism of QM has been supplied in the papers quoted above according to which all physical properties are objective for every state of the physical system that is considered.  The new interpretation adopts a purely semantic version of objectivity (therefore it has been called \emph{Semantic Realism}, or \emph{SR}, \emph{interpretation}) and entails that QM is semantically incomplete. Unfortunately, it is scarcely intuitive and may seem founded on a problematic epistemological analysis to many pragmatically oriented physicists. Therefore the same authors have proposed  an \emph{SR model} \cite{ga02,gp04} and an \emph{extended SR} (briefly, \emph{ESR}) \emph{model} \cite{gp04,ga03,ga05,gps06} which introduce a more intuitive picture of the physical world which is consistent with the SR interpretation. We intend to show in this paper that the ESR model provides a general scheme for an objective theory embodying the mathematical apparatus of QM. Hence we present in Sect. \ref{modello} a revised and improved version of this model, stressing that it implies a reinterpretation of quantum probabilities which has long--ranging implications. Then we show in Sect. \ref{consistenza} that objectivity of (macroscopic) physical properties holds in the ESR model if one adds a further reasonable axiom on probabilities, and argue that the predictions of the ESR model differ from those following from the standard interpretation of QM, which makes it possible, in principle, to check which interpretation is correct. In particular, we consider in Sect. \ref{disgeneralizzate} the physical situation introduced in the literature to obtain the \emph{standard Bell--Clauser--Horne--Shimony--Holt} (\emph{BCHSH}) \emph{inequality} \cite{chsh69} from the point of view of the ESR model, and prove that, whenever idealized measurements are performed and all prepared physical objects are taken into account, a \emph{modified BCHSH inequality} holds. Then we discuss in Sect. \ref{modifiedBCHSH} how the quantum predictions, which refer to the set of all physical objects that are detected, can be inserted in the modified BCHSH inequality, and which kind of experimental tests can be performed. Furthermore we show in Sect. \ref{unfairsampling} that a \emph{standard BCHSH inequality} holds at a microscopic (purely theoretical) level, hence we conclude that the standard BCHSH inequality, the modified BCHSH inequality and the quantum predictions hold together in the ESR model because they refer to different parts of the picture provided by the model. This is in our opinion a relevant result, because it shows how a long lasting conflict among different inequalities can be settled,\footnote{We stress that the ESR model is deeply different from the approaches that try to vindicate \emph{local realism} by questioning the interpretation of the experimental results obtained till now because of the low efficiencies of the detectors. Indeed these approaches consider standard Bell's and quantum inequalities as competing theoretical results which cannot be decided on the basis of the available experimental data, while the ESR model reconciles these inequalities. The interested reader can find a more detailed discussion of this topic in \cite{g07}.} hence we also provide in Sect. \ref{unfairsampling} an intuitive explanation of this settlement in terms of an unconventional kind of unfair sampling.

\section{The ESR model\label{modello}}
As we have anticipated in Sect. \ref{intro}, the ESR model has been proposed by one of the authors few years ago \cite{ga02,gp04,ga03,ga05,gps06} and we aim to present here an improved version of it, together with a more complete treatment of idealized measurements.

The basic notions of the ESR model can be divided in three groups.

(i) Standard primitive and derived notions: \emph{physical system}, \emph{preparing device}, \emph{state}, \emph{physical object}. In particular, a state of a physical system $\Omega$ is defined as a class of physically equivalent preparing devices \cite{bc81,l83}. A physical object is defined as an individual example $x$ of $\Omega$, obtained by activating a preparing device $\pi$, and we say that ``$x$ is (prepared) in the state $S$'' if $\pi \in S$.

(ii) New observational entities: \emph{generalized observables}. Every physical system $\Omega$ is associated with a set $\mathcal O$ of generalized observables. Every generalized observable $A_0$ (here meant as a class of physically equivalent measuring apparatuses, without any reference to a mathematical representation) is obtained in the ESR model by considering an observable $A$ of QM with set of possible values $\Xi$ on the real line $\Re$ and adding a further outcome $a_0$ (\emph{no--registration outcome} of $A_0$) that does not belong to $\Xi$, so that the set of possible values of $A_0$ is $\Xi_0=\{ a_0 \}  \cup \Xi$.\footnote{If $\Xi=\Re$ the observable $A$ can be substituted, without loss of generality, by the observable $f(A)$, with $f$ a bijective mapping which maps $\Re$ onto a proper subset, say $\Re^{+}$, of $\Re$.\label{borel}}

(iii) New theoretical entities: \emph{microscopic properties}. Every physical system $\Omega$ is characterized by a set $\mathcal E$ of microscopic properties. For every physical object $x$, the set $\mathcal E$ is partitioned in two classes, the class of properties that are possessed by $x$ and the class of properties that are not possessed by $x$, independently of any measurement procedure. We stress that different physical objects in the same state $S$ may possess different microscopic properties (but assigning the state $S$ of $x$ imposes some limits on the subset of microscopic properties that can be possessed by $x$, see footnote \ref{certrue}). 

Let ${\mathbb B}(\Re)$ be the $\sigma$--algebra of all Borel subsets of $\Re$. The introduction of generalized observables allows us to define the set ${\mathcal F}_{0}$ of all \emph{macroscopic properties} of $\Omega$,
\begin{equation}
{\mathcal F}_{0} \ =  \ \{ (A_0, X) \ | \  A_0 \in {\mathcal O}, \ X \in {\mathbb B}(\Re) \},
\end{equation}
and the set ${\mathcal F} \subset {\mathcal F}_{0}$ of all macroscopic properties associated with observables of QM,
\begin{equation} 
{\mathcal F} \ = \ \{ (A_0, X) \ | \  A_0  \in {\mathcal O}, \ X \in {\mathbb B}(\Re), \ a_0 \notin X \}.
\end{equation}
For every ${A}_{0} \in {\mathcal O}$, different Borel sets containing the same subset of $\Xi_{0}$ define physically equivalent properties. For the sake of simplicity we convene that, whenever we mention macroscopic properties in the following, we actually understand such classes of physically equivalent macroscopic properties. Furthermore, we agree to write simply \emph{observable} in place of \emph{generalized observable} whenever no misunderstanding is possible.

We establish a link between microscopic properties of $\mathcal E$ and macroscopic properties of $\mathcal F$ by means of the following assumption.

\emph{Ax. 1.} \emph{A bijective mapping $\varphi: {\mathcal E} \longrightarrow {\mathcal F} \subset {\mathcal F}_{0}$ exists}. 

Let us describe now an \emph{idealized} measurement of a macroscopic property $F=(A_0, X)$ on a physical object $x$ in the state $S$. We assume that such a measurement consists of a registration, performed by means of a dichotomic registering device (which may be constructed by using one of the apparatuses associated with $A_0$), whose outcomes we denote by \emph{yes} and \emph{no}. The measurement yields outcome yes (no) if the value of $A_0$ belongs (does not belong) to $X$, and we say in this case that $x$ \emph{displays} (\emph{does not display}) $F$.\footnote{The introduction of microscopic properties implies that the term ``possessed'' and ``not possessed'' may be misleading if referred to macroscopic properties in our framework. Hence we use them only with reference to microscopic properties and introduce a new terminology to describe the relations between macroscopic properties and physical objects.} It is then important to note that $x$ displays (does not display) $F$ if it does not display (displays) $F^{c}=(A_0, \Re \setminus X)$. Now, we assume that the set of all microscopic properties possessed by $x$ induces a probability (which is either 0 or 1 if the model is \emph{deterministic}) that the apparatus react (\emph{detection probability}). Whenever the apparatus reacts, $x$ displays $F$ if it possesses the microscopic property $f=\varphi^{-1}((A_0, X\setminus \{ a_0 \}))$ (where $(A_0, X\setminus \{ a_0 \})$ coincides with F iff $F \in \mathcal F$), otherwise it displays $F^{c}$. Whenever the apparatus does not react, $x$ displays $F$ if $F \in {\mathcal F}_{0} \setminus {\mathcal F}$, while it displays $F^{c}$ if $F \in \mathcal F$.

The above description implies that the microscopic properties determine the probability of an outcome (or the outcome itself if the model is deterministic), which therefore does not depend on features of the measuring apparatus (flaws, termal noise, etc.) nor is influenced by the environment. In this sense idealized measurements are ``perfectly efficient'', and must be considered as a limit of concrete measurements in which the specific features of apparatuses and environment must instead be taken into account.

We must still place properly quantum probability in our picture. To this end, let us suppose that the device $\pi \in S$ is activated repeatedly, hence a finite set $\mathscr S$ of physical objects in the state $S$ is prepared. Then, $\mathscr S$ can be partitioned into subsets ${\mathscr S}^{1}, {\mathscr S}^{2}, \ldots, {\mathscr S}^{n}$ such that each subset collects all objects possessing the same microscopic properties. We briefly say that the objects in ${\mathscr S}^{i}$ ($i=1,2, \ldots, n$) are in some \emph{microscopic state} $S^{i}$. This suggests us to associate every state $S$ with a family of microscopic states $S^{1}, S^{2}, \ldots$ and characterize $S^{i}$ ($i=1,2, \ldots$) by the set of all microscopic properties that are possessed by any physical object in ${S}^{i}$ (hence also microscopic states play the role of theoretical entities in the ESR model). Let us now consider a physical object $x$ in $S^{i}$, and let us suppose that a measurement of a macroscopic property $F=(A_0, X) \in \mathcal F$ is performed on it. It follows from our description of the measurement process that, whenever $x$ is detected, $x$ displays $F$ if and only if the microscopic property $f=\varphi^{-1}(F)$ is one of the microscopic properties characterizing $S^{i}$. 
We are thus led to introduce the following probabilities.

${p}_{S}^{i,d}(F)$: the probability that $x$ be detected when $F$ is measured on it.

${p}_{S}^{i}(F)$: the conditional probability that $x$ display $F$ when it is detected (which is 0 or 1 since $x$ either possesses $\varphi^{-1}(F)$ or not).

${p}_{S}^{i,t}(F)$: the joint probability that $x$ be detected and display $F$.

Hence, we get
\begin{equation} \label{formuladipartenza_i}
{p}_{S}^{i,t}(F)={p}_{S}^{i,d}(F) {p}_{S}^{i}(F).
\end{equation}
Eq. (\ref{formuladipartenza_i}) is purely theoretical, since one can never know if a physical object is in the microstate $S^i$. Therefore, let us consider a physical object in the state $S$ and introduce a further conditional probability, as follows.

${p}(S^{i}|S)$: the conditional probability that $x$, which is in the macroscopic state $S$, be in the microstate $S^i$.

The joint probability that $x$ be in the state $S^i$, be detected and display $F$ is thus given by ${p}(S^{i}|S) {p}_{S}^{i,t}(F)$. Hence the overall probability ${p}_{S}^{t}(F)$ that $x$ be detected and display $F$ is given by
\begin{equation} \label{formuladipartenza_t}
{p}_{S}^{t}(F)=\sum_{i}{p}(S^{i}|S){p}^{i,t}_{S}(F).
\end{equation}
Moreover, the probability ${p}^{d}_{S}(F)$ that $x$ be detected when $F$ is measured is given by
\begin{equation} \label{formuladipartenza_d}
{p}_{S}^{d}(F)=\sum_{i}{p}(S^{i}|S){p}^{i,d}_{S}(F).
\end{equation}
Let us define now
\begin{equation} \label{formuladipartenza_qm}
{p}_{S}(F)=\frac{\sum_{i}{p}(S^{i}|S){p}^{i,t}_{S}(F)}{\sum_{i}{p}(S^{i}|S){p}^{i,d}_{S}(F)}.
\end{equation}
Then, we get
\begin{equation} \label{formuladipartenza}
{p}_{S}^{t}(F)={p}_{S}^{d}(F){p}_{S}(F).
\end{equation} 
Eq. (\ref{formuladipartenza}) is the fundamental equation of the ESR model. Let us therefore discuss the two factors that appear in it.

Let us begin with the \emph{detection probability} ${p}_{S}^{d}(F)$. We have seen that, since we are dealing here with idealized measurements, the occurrence of the outcome $a_0$ is attributed only to the set of microscopic properties possessed by $x$, which determines the probability ${p}_{S}^{i,d}(F)$. Hence, ${p}_{S}^{i,d}(F)$ neither depends on features of the measuring apparatus nor is influenced by the environment. Furthermore, the conditional probability ${p}(S^{i}|S)$ depends only on $S$. Therefore, Eq. (\ref{formuladipartenza_d}) implies that ${p}_{S}^{d}(F)$ depends only on the microscopic properties of the physical objects in $S$.

Let us come to ${p}_{S}(F)$. By using Eqs. (\ref{formuladipartenza_i}) and (\ref{formuladipartenza_qm}) we get $0 \le {p}_{S}(F) \le 1$. Moreover, the interpretations of ${p}_{S}^{t}(F)$ and ${p}_{S}^{d}(F)$ in Eq. (\ref{formuladipartenza}) show that ${p}_{S}(F)$ can be interpreted as the conditional probability that a physical object $x$ display $F$ when it is detected. This interpretation of the term ${p}_{S}(F)$ in Eq. (\ref{formuladipartenza}) provides a basis for the introduction of the main assumption of the ESR model. 

\emph{Ax. 2.} \emph{If $S$ is a pure state, the probability ${p}_{S}(F)$ can be evaluated by using the same rules that yield the probability of $F$ in the state $S$ according to QM}.

Ax. 2 implies a new interpretation of the probabilities provided by standard quantum rules, which are now regarded as conditional rather than absolute.\footnote{Note that, if the state $S$ is assigned and the property $F$ is such that ${p}_{S}(F)=1$, every physical object that is detected necessarily possesses the microscopic property $f=\varphi^{-1}(F)$. This specifies the limits mentioned in (iii). \label{certrue}} The old and the new interpretation of quantum probabilities coincide if ${p}_{S}^{d}(F)=1$ for every state $S$ and property $F$. If there are states and properties such that ${p}_{S}^{d}(F)<1$, instead, the difference between the two interpretations is conceptually relevant. To better grasp this difference, let us regard probabilities as large number limits of frequencies in ensembles of physical objects,\footnote{We adopt this na\"\i ve interpretation of physical probabilities here for the sake of simplicity. A more sophisticated treatment would associate quantum measurements with random variables, require that distribution functions approach experimental frequencies, etc. Our conclusions, however, would not be modified by the adoption of this more general and rigorous machinery.} and let us consider ensembles of physical objects in the state $S$. Then, ${p}_{S}(F)$ is the limit of the ratio between the number of objects in a given ensemble that are detected and display the property $F$, and the number of objects that are detected. The same quantity would be interpreted in QM as the limit of the ratio between the number of objects in a given ensemble that display the property $F$ and the number of objects in the ensemble. Hence, the ESR model introduces a non--orthodox interpretation of quantum probabilities. But it must be stressed that Ax. 2 implies that, as far as the probability $p_{S}(F)$ is concerned, the physical system $\Omega$ can be associated with a Hilbert space $\mathscr H$ and its pure states and macroscopic properties in $\mathcal F$ can be represented by means of (unit) vectors on $\mathscr H$ and (orthogonal) projection operators on $\mathscr H$, respectively. 

Let us complete our discussion by considering a macroscopic property $G=(A_0,Y) \in {\mathcal F}_{0}\setminus {\mathcal F}$. In this case we can introduce the macroscopic property $F=(A_0, Y \setminus \{ a_0 \}) \in {\mathcal F}$ and get, because of our description of the measurement procedure,
\begin{equation} \label{formuladipartenza_F0^i}
{p}_{S}^{i,t}(G)=(1-{p}_{S}^{i,d}(F))+{p}_{S}^{i,t}(F)
\end{equation} 
where $p_{S}^{i,t}(G)$ is the probability that $x$ display the property $G$ whenever it is in the microscopic state $S^{i}$. By using Eqs. (\ref{formuladipartenza_t})--(\ref{formuladipartenza}) we then obtain
\begin{equation} \label{formuladipartenza_F0}
{p}_{S}^{t}(G)=(1-{p}_{S}^{d}(F))+{p}_{S}^{t}(F)=1-{p}_{S}^{d}(F)(1-{p}_{S}(F))
\end{equation} 
where $p_{S}^{t}(G)$ is the overall probability that $x$ display $G$ and $p_{S}(F)$ can be evaluated by using standard quantum rules. 

We conclude this section by observing that every microscopic property $f$ can be associated with a dichotomic \emph{hidden variable}, which takes value 1 (0) if $f$ is possessed (not possessed) by the physical object $x$ that is considered. Equivalently, a microscopic state can be seen as the value of a hidden variable $\lambda$ specifying all microscopic properties of $x$. Hence, the ESR model provides a general scheme for a hidden--variables theory which exhibits some similarities with existing hidden--variables theories or models, but is different from all previous proposals because of its reinterpretation of quantum probabilities. Besides, it must be stressed that no hidden variable associated with the measuring apparatuses occurs in this theory because the ESR model is noncontextual.

\section{Objectivity and physical predictions in the ESR model \label{consistenza}}
As we have anticipated in Sect. \ref{intro}, we intend to show in this section that the ESR model provides, by introducing microscopic properties, an intuitive (set--theoretical) picture of the physical world in which macroscopic properties can be considered objective. This picture allows one to make physical predictions even if some theoretical entities in the ESR model have no mathematical representation at this stage.

First of all, let us adopt the following definition of objectivity (which particularizes the definition of objectivity introduced in Sect. \ref{intro}) when referring to hidden--variables theories.

\emph{Objectivity}. A hidden--variables theory is objective iff for every macroscopic physical property $F$ and state $S$ the outcome of a measurement of $F$ on a physical object $x$ in the state $S$ is determined only by hidden variables associated with $x$.

The above notion of objectivity is strictly linked with the notion of \emph{contextuality}, which is broadly used in the literature. To be precise, the latter notion occurs in at least two different senses \cite{gps06}.

(i) \emph{Contextuality$_1$}. A hidden--variables theory is contextual if the outcome of a measurement of a (macroscopic) property $F$ on a physical object $x$ may depend on hidden variables associated with the set of (compatible) measurements that are simultaneously performed on $x$, not only on hidden variables associated with $x$ (standard notion, see, \emph{e.g.}, \cite{ks67,m93}).

(ii) \emph{Contextuality$_2$}. A hidden--variables theory is contextual if the outcome of the measurement of a (macroscopic) property $F$ on a physical object $x$ may depend on hidden variables associated with the specific registration that is used to perform the measurement, not only on hidden variables associated with $x$ (\emph{hidden measurement approach}, see, \emph{e.g.}, \cite{aa04}, and \emph{probabilistic opposition}, see, \emph{e.g.}, \cite{k07}).

The two kinds of contextuality may coexist and both imply nonobjectivity. Conversely, objectivity implies noncontextuality$_1$ and noncontextuality$_2$. Hence we can use the word \emph{objectivity} as a synonym of \emph{noncontextuality}, without specifying whether contextuality$_1$ or contextuality$_2$ is understood. 

Let us consider now Eq. (\ref{formuladipartenza_i}). Because of the definition of ${p}_{S}^{i}(F)$, and because ${p}_{S}^{i,d}(F)$ only depends on microscopic properties, this equation shows that the probability ${p}_{S}^{i,t}(F)$ is completely determined by the value $S^{i}$ of the hidden variable (equivalently, by the set of all microscopic properties possessed by a physical object $x$ in $S^{i}$), hence it is independent of the measurement context, that is, it is noncontextual. But, of course, noncontextuality of probabilities (which holds also in QM, see, \emph{e.g.}, \cite{m93}) does not imply objectivity of macroscopic physical properties. Nevertheless, whenever the ESR model is deterministic, the microscopic properties actually determine the outcome of a measurement of $F$ (Sect. \ref{modello}), hence objectivity of macroscopic properties holds in this case. If the ESR model, instead, is nondeterministic, we cannot deduce objectivity but can introduce a new assumption which implies it.

\emph{Ax. 3.} \emph{For every microscopic state $S^{i}$, the probability ${p}_{S}^{i,d}(F)$ admits an} epistemic (\emph{or} ignorance) \emph{interpretation} (see, \emph{e.g.}, \cite{bc81}) \emph{in terms of further unknown features of the physical objects in the state $S^{i}$}.

Indeed, Ax. 3 implies that a parameter $\mu$ exists which determines, together with $S^{i}$, whether the physical object $x$ is detected whenever the property $F$ is measured on it, \emph{i.e.}, $\mu$ and $S^{i}$ determine whether the outcome $a_0$ occurs or not (note that $\mu$ can be interpreted as denoting a subset of further microscopic properties possessed by $x$, selected in a new set of microscopic properties that do not correspond to macroscopic properties via $\varphi$). Since $S^{i}$ determines all macroscopic properties of $x$ whenever $x$ is detected, all macroscopic properties are determined by the pair $(\mu, S^{i})$, hence the ESR model is objective in the sense specified above. 

Because of Ax. 3 and our remark at the end of Sect. \ref{modello} the ESR model can be considered as a scheme for a noncontextual hidden--variables theory with reinterpretation of quantum probabilities. But, of course, it is not yet a general theory. Indeed, one should still introduce mathematical representations of generalized observables and macroscopic properties, rules for evaluating the detection probabilities, and evolution laws (first steps in this direction are proposed in \cite{sg08,gs09a}). From the point of view of the ESR model one needs a completion of this kind, in particular, to deal with measurements in terms of interactions between macroscopic apparatuses and microscopic physical objects, which explains the failure of the attempts at providing an exhaustive theory of quantum measurements in QM.\footnote{Note that if the ESR model is nondeterministic and one avoids introducing Ax. 3, contextuality cannot be excluded. In this case one can assume that the outcome of a measurement of a macroscopic property is determined by a pair $(\nu, S^{i})$, where $\nu$ denotes a hidden variable (or a set of hidden variables) associated with the measurement context (\emph{contextual ESR} model). The contextual ESR model provides a scheme for a general hidden--variables theory which embodies the quantum formalism and reinterprets quantum probabilities but preserves contextuality. It is then interesting to observe that this theory may be \emph{local} or \emph{nonlocal}, depending on the assumptions on the parameter $\nu$. In the former case it converges with the proposals of the probabilistic opposition \cite{k07}, while in the latter case it converges with the perspective of the hidden measurement approach \cite{aa04}.}

Leaving apart the above general problems and limiting ourselves to the ESR model as presented here, let us consider the statistical predictions that can already be obtained. Let a set of idealized measurements be performed on an ensemble of physical objects in a state $S$ (we recall that the ESR model deals with idealized measurements only, hence the lack of efficiency of actual measuring apparatuses must be taken into account separately). It follows from the general features of the ESR model (in particular, Ax. 2) that its predictions can be partitioned in two classes.

(a) Predictions concerning the subensemble of all physical objects that are detected by the measurements. They are obtained by using the quantum formalism (see Ax. 2), hence formally coincide with the predictions of QM, but QM would interpret them as referring to the whole ensemble. 

(b) Predictions concerning the whole ensemble. Here the detection probability ${p}_{S}^{d}(F)$ plays an essential role, but one may think at first sight that these predictions can be only qualitative because of the lack of a general theory for ${p}_{S}^{d}(F)$. On the contrary, by considering a special case we show in Sect. \ref{modifiedBCHSH} that the assumptions of the ESR model (in particular, objectivity) imply quantitative conditions on the possible values of $p_{S}^{d}(F)$ which entail that $p_{S}^{d}(F)<1$ for some $S \in \mathcal S$ and $F \in \mathcal F$. 

It follows from (a) and (b) that the predictions of the ESR model are essentially different from those of QM. Moreover, one can try to get empirical information about $p_{S}^{d}(F)$ from experimental data and compare it with the conditions imposed by the ESR model (see in particular Sect. \ref{modifiedBCHSH}). Hence, at least in principle, the ESR model is \emph{falsifiable}.\footnote{In practice it may be hard to check the ESR model in this way. Indeed, if the intrinsic lack of efficiency of any measuring device is schematized by multiplying ${p}_{S}^{d}(F)$ by a factor $k$, with $0 \le k \le 1$, it may be difficult to distinguish empirically $k$ from ${p}_{S}^{d}(F)$. \label{cappa}}

\section{The modified BCHSH inequality\label{disgeneralizzate}}
The ESR model presented in Sect. \ref{modello} introduces several theoretical entities (microscopic properties and states) that are not operationally defined. But these entities do not appear in Eq. (\ref{formuladipartenza}), which can be postulated \emph{a priori} if one wants to reinterpret quantum probabilities without introducing underlying models. Basing only on Eq. (\ref{formuladipartenza}) we intend to discuss in this section some consequences of objectivity in the ESR model by reconsidering the physical situations that led to the standard BCHSH inequality.

First of all, let us observe that objectivity as defined in Sect. \ref{consistenza} implies \emph{local realism} in the standard sense in the literature \cite{b64,chsh69,epr35},\footnote{The use of the phrase ``local realism'' in the context of Bell's Theorem has recently been disputed \cite{n07}. We add that the meaning of the word ``realism'' in this phrase does not match the meaning of the word ``realistic'' used by Busch \emph{et al.} \cite{blm91} and mentioned in Sect. \ref{intro}. Nevertheless we will not break with the standard language here because definition R is widely used in the literature and has a precise meaning when translated in terms of hidden variables, even thought it does not fit in well with traditional notions of realism adopted by philosophers.} \emph{i.e.}, the join of the assumptions of \emph{realism},

\emph{R}: \emph{the values of all observables of a physical system in a given state are predetermined for any measurement context},

\noindent
and \emph{locality},

\emph{LOC}: \emph{if measurements are made at places remote from one another on parts of a physical system which no longer interact, the specific features of one of the measurements do not influence the results obtained with the others}.

It seems at first sight that the standard BCHSH inequality must then hold in the ESR model. But such a conclusion is false. Indeed the proof of this inequality does not rest only on R and LOC, but also on the implicit condition that in an ideal measurement all physical objects on which the measurement is performed are detected. This condition is not fulfilled by the idealized measurements introduced in the ESR model, where the set of possible outcomes of a generalized observable contains $a_0$. Hence, we can prove that \emph{modified} instead of standard BCHSH inequalities hold in the ESR model whenever all physical objects that are prepared are taken into account.

To begin with, let us resume the physical situation introduced in the literature to obtain a standard BCHSH inequality.

Let $\Omega$ be a compound physical system made up of two far away subsystems $\Omega_1$ and $\Omega_2$, and let $A({\bf a})$ and $B({\bf b})$ be dichotomic observables  of $\Omega_1$ and $\Omega_2$, respectively, depending on the parameters ${\bf a}$ and ${\bf b}$ and taking either value $-1$ or $+1$. Then, the standard treatment assumes R and LOC and defines a correlation function
\begin{equation} \label{lhvt}
P({\bf a}, {\bf b})= \int_{\Lambda} d\lambda \rho(\lambda) A(\lambda, {\bf a}) B(\lambda, {\bf b}),
\end{equation}
where $\lambda$ is a deterministic \emph{hidden variable} whose value ranges over a domain $\Lambda$ when measurements on different examples of $\Omega$ in a given state $S$ are considered, $\rho(\lambda)$ is a probability distribution on $\Lambda$, $A(\lambda, {\bf a})$ and $B(\lambda, {\bf b})$ are the values of $A({\bf a})$ and $B({\bf b})$, respectively. By using Eq. (\ref{lhvt}) one gets the \emph{standard BCHSH inequality}
\begin{equation} \label{chsh_69}
|P({\bf a}, {\bf b})-P( {\bf a}, {\bf b'})|+|P({\bf a'}, {\bf b})+P({\bf a'}, {\bf b'})| \le 2.
\end{equation}
As we have seen at the beginning of this section, however, the proof of Eq. (\ref{chsh_69}) also rests on the condition, which is usually left implicit, that in a (ideal) measurement all physical objects on which the measurement is performed be detected. This condition is not fulfilled by the idealized measurements of the ESR model, where the no--registration outcome occurs in every generalized observable that is considered. Therefore the dichotomic observables $A({\bf a})$, $B({\bf b})$, $A({\bf a'})$ and $B({\bf b'})$ must be substituted by the trichotomic observables $A_0({\bf a})$, $B_0({\bf b})$, $A_0({\bf a'})$ and $B_0({\bf b'})$, respectively, in each of which a no--registration outcome is added to the outcomes $+1$ and $-1$. Let us agree to consider trichotomic observables such that all no--registration outcomes coincide with $0$.\footnote{Note that, if $A_0$ is a generalized observable obtained from an observable $A$ of QM by introducing the no--registration outcome $a_0 \ne 0$, one can easily construct a new observable such that its no--registration outcome is $0$. Indeed, one can introduce a Borel function $\chi$ on $\Re$ which is bijective on $\Xi_0$ and such that $\chi(a_0)=0$, consider the observable $\chi(A)$ of QM, add the outcome $0$ to its set of possible values, and get a generalized observable $\chi(A_0)$, with ${p}_{S}^{d}(\chi(A_0))={p}_{S}^{d}(A_0)$.\label{legittima}} Then we recall that the range of values of the hidden variable $\lambda$ is the subset of all microscopic states associated with the macroscopic state $S$ (see Sect. \ref{modello}). Hence the correlation function in Eq. (\ref{lhvt}) must be substituted by a \emph{generalized correlation function}
\begin{equation} \label{generalized_corr_func}
P(A_0({\bf a}),B_0({\bf b}))=\sum_{i} {p}(S^{i}|S)A_0(S^i, {\bf a}) B_0(S^i, {\bf b}) ,
\end{equation}
where $A_0(S^i, {\bf a})$ and $B_0(S^i, {\bf b})$ denote the values of $A_0({\bf a})$ and $B_0({\bf b})$, respectively, when the hidden variable takes value $S^{i}$ (which implies that we must consider an ESR model which is deterministic in the sense explained in Sect. \ref{modello}). We can now follow the standard procedures leading to Eq. (\ref{chsh_69}). Since $|A_0(S^i, {\bf a})|\le 1$ we get
\begin{equation}
|P(A_0({\bf a}),B_0({\bf b}))-P(A_0({\bf a}),B_0({\bf b'}))| \le \sum_{i} {p}(S^i|S) |B_0(S^i, {\bf b})-B_0(S^i, {\bf b'})|
\end{equation}
and, similarly,
\begin{equation}
|P(A_0({\bf a'}),B_0({\bf b}))+P(A_0({\bf a'}),B_0({\bf b'}))| \le \sum_{i} {p}(S^i|S) |B_0(S^i, {\bf b})+B_0(S^i, {\bf b'})|.
\end{equation}
Now, we have
\begin{equation}
|B_0(S^i, {\bf b})-B_0(S^i, {\bf b'})|+|B_0(S^i, {\bf b})+B_0(S^i, {\bf b'})| \le 2
\end{equation}
and
\begin{equation}
\sum_{i} {p}(S^i|S)=1,
\end{equation}
hence we obtain the \emph{modified BCHSH inequality}
%\begin{eqnarray}
\begin{equation}
|P(A_0({\bf a}), B_0({\bf b}))-P(A_0({\bf a}), B_0({\bf b'}))| 
%\nonumber \\
+ |P(A_0({\bf a'}), B_0({\bf b}))+ P(A_0({\bf a'}), B_0({\bf b'}))|\le 2 , \label{chsh_gs}
\end{equation}
%\end{eqnarray}
which replaces Eq. (\ref{chsh_69}) in the ESR model.

\section{Modified BCHSH inequality and quantum predictions\label{modifiedBCHSH}}
The generalized correlation function in Eq. (\ref{generalized_corr_func}), hence the modified BCHSH inequality, clearly refer (via the conditional probability $p(S^{i}|S)$ that a physical object in a state $S$ be in the microstate $S^{i}$) to the set of all objects that are prepared in the state $S$, just as the correlation function in Eq. (\ref{lhvt}). In this section we want to insert in Eq. (\ref{generalized_corr_func}) the predictions that can be attained by using the rules of QM, which refer to the subset of all detected physical objects only. This can be done in general by using the fundamental equation of the ESR model, but the resulting formulas are rather complicate and their interpretation is not immediate. To better highlight the peculiarities of the ESR approach we therefore consider a special case in which some simplificative conditions hold, as follows.

First of all, let us agree to consider discrete observables and pure states only. Hence the set $\Xi_0$ of all possible values of a generalized observable $A_0$ will be given by $\Xi_0=\{ a_0 \} \cup \{ a_1, a_2, \ldots \}$, where $\Xi=\{ a_1, a_2, \ldots \}$ is the set of all possible values of the observable $A$ of QM from which $A_0$ is obtained. Measuring $A_0$ is thus equivalent to measuring the properties $F_0= (A_0, \{ a_0 \})$, $F_1= (A_0, \{ a_1 \})$, $F_2= (A_0, \{ a_2 \})$, \ldots simultaneously, hence, for every $F_n$ such that $n \in {\mathbb N}$, Eq. (\ref{formuladipartenza}) holds with $F_n$ in place of $F$, and we briefly write $a_n$ instead of $F_n$ in it. Then, let us agree to take into account only (discrete) observables that satisfy the following simplificative condition:

(i) the detection probability depends on the observable but not on its specific value. 

Because of condition (i), we write ${p}_{S}^{d}(A_0)$ instead of ${p}_{S}^{d}(a_n)$ and get from Eqs. (\ref{formuladipartenza}) and (\ref{formuladipartenza_F0})
\begin{equation} \label{a_n}
{p}_{S}^{t}(a_n)={p}_{S}^{d}(A_0){p}_{S}(a_n).
\end{equation}  
and
\begin{equation} \label{a_0}
{p}_{S}^{t}(a_0)=1 - {p}_{S}^{d}(A_0),
\end{equation}  
respectively. Eqs. (\ref{a_n}) and (\ref{a_0}) can be used to evaluate the expectation value ${\langle A_0 \rangle}_{S}$ of $A_0$ in the state $S$,
\begin{equation} \label{expectation_value}
{\langle A_0 \rangle}_{S}=a_0 (1 - {p}_{S}^{d}(A_0))+ {p}_{S}^{d}(A_0) {\langle A \rangle}_{S}, 
\end{equation}
where 
\begin{equation}
{\langle A \rangle}_{S}=\sum_{n}a_n {p}_{S}(a_n)
\end{equation}
is the \emph{conditional expectation value} of $A_0$, which coincides with the standard quantum expectation value of the observable $A$ of QM from which $A_0$ is obtained, but has a different physical interpretation, because it represents the mean value of $A_0$ whenever only detected physical objects are taken into account. 

Let us assume further that for each observable considered in the following an idealized measurement exists which satisfies the following conditions:

(ii) the measurement may change the state of the physical object $x$ on which it is performed but it does not destroy $x$, even if $x$ is not detected;

(iii) if the physical object $x$ is detected and a given outcome is obtained, the state of $x$ after the measurement can be predicted by using the Hilbert space representation of pure states and the projection postulate of QM;

(iv) the measurement is minimally perturbing, in the sense that the state of a physical object $x$ is not changed by the measurement whenever $x$ is not detected.

Then, let us consider a compound physical system $\Omega$ made up by two far apart subsystems $\Omega_1$ and $\Omega_2$, and let $A_0, B_0$ be observables of the component subsystems $\Omega_1$ and $\Omega_2$, respectively. In this case objectivity of properties implies that the change of the state of $\Omega$ induced by a measurement of $A_0$ on $\Omega_1$ must not affect the detection probability associated with a simultaneous measurement of $B_0$ on $\Omega_2$ (whose possible outcomes will be denoted by $b_0, b_1, b_2, \ldots$). By using this remark and conditions (i)--(iv) we can calculate the probabilities ${p}_{S}^{t}(a_n,b_p)$, ${p}_{S}^{t}(a_n,b_0)$, ${p}_{S}^{t}(a_0,b_p)$, ${p}_{S}^{t}(a_0,b_0)$ (with $n,p \in {\mathbb N}$) of obtaining the pairs of outcomes $(a_n,b_p)$, $(a_n,b_0)$, $(a_0,b_p)$, $(a_0,b_0)$, respectively, in simultaneous measurements of ${A}_0$ and ${B}_0$ on a physical object $x$ in the state $S$. We get    
\begin{equation} \label{assnp}
{p}_{S}^{t}(a_n,b_p)={p}_{S}^{d}(A_0){p}_{S}^{d}(B_0){p}_{S}(a_n,b_p),
\end{equation}  
\begin{equation}
{p}_{S}^{t}(a_n,b_0)={p}_{S}^{d}(A_0)(1-{p}_{S}^{d}(B_0)) {p}_{S}(a_n),
\end{equation}
\begin{equation}
{p}_{S}^{t}(a_0,b_p)=(1-{p}_{S}^{d}(A_0)){p}_{S}^{d}(B_0){p}_{S}(b_p),
\end{equation}
\begin{equation} \label{ass00}
{p}_{S}^{t}(a_0,b_0)=(1-{p}_{S}^{d}(A_0))(1-{p}_{S}^{d}(B_0)),
\end{equation}
where $p_{S}(a_n,b_p)$ is the quantum probability of obtaining the pair $(a_n,b_p)$ when measuring $A_0$ and $B_0$. Note that Eqs. (\ref{assnp})--(\ref{ass00}) contain only two probabilities that cannot be evaluated by using the rules of QM, that is, ${p}_{S}^{d}(A_0)$ and ${p}_{S}^{d}(B_0)$. 

Let us define now a \emph{generalized correlation function} $P(A_0,B_0)$, as follows
\begin{eqnarray} 
P(A_0,B_0)=\sum_{n,p} a_n b_p {p}_{S}^{t}(a_n,b_p)+\sum_{n} a_n b_0 {p}_{S}^{t}(a_n,b_0) \nonumber \\
+ \sum_{p} a_0 b_p {p}_{S}^{t}(a_0,b_p)+ a_0 b_0 {p}_{S}^{t}(a_0,b_0). \label{corrfunc}  
\end{eqnarray}
By using Eqs. (\ref{assnp})--(\ref{ass00}) and introducing the further assumption:

(v) $a_0=0=b_0$

\noindent
(hence, for every $n,p \in {\mathbb N}$, $a_n \ne 0 \ne b_p$) we get that the generalized correlation function, for the special class of observables that we are considering, is given by
\begin{equation} \label{corrfunc_ass}
P(A_0,B_0)={p}_{S}^{d}(A_0) {p}_{S}^{d}(B_0) {\langle AB \rangle}_{S},
\end{equation}
where
\begin{equation} \label{corrfunc_QM}
{\langle AB \rangle}_{S}=\sum_{n,p} a_n b_p {p}_{S}(a_n,b_p).
\end{equation}
Because of the interpretation of $p_{S}(a_n,b_p)$, ${\langle AB \rangle}_{S}$ formally coincides with the standard quantum expectation value in the state $S$ of the product of the (compatible) observables $A$ and $B$ from which $A_0$ and $B_0$, respectively, are obtained. But its physical interpretation is different, for it represents a mean value whenever only detected physical objects are taken into account. As the standard correlation function in the literature, $P(A_0,B_0)$ may provide an index of the correlation among the outcomes of $A_0$ and $B_0$ in the state $S$. 

Let us come now to the modified BCHSH inequality, and let us recall that each trichotomic observable that occurs in it has a no--registration outcome which is equal to $0$, so that it satisfies condition (v). Let us further suppose that it also satisfies conditions (i)--(iv). Then we can use Eq. (\ref{corrfunc_ass}) and get from Eq. (\ref{chsh_gs}) \cite{s07}
\begin{eqnarray}
{p}_{S}^{d}(A_0({\bf a}))|{p}_{S}^{d}(B_0({\bf b})) {\langle A({\bf a})B({\bf b})\rangle}_{S}- {p}_{S}^{d}(B_0({\bf b'})){\langle A({\bf a})B({\bf b'})\rangle}_{S} | \nonumber \\
+{p}_{S}^{d}(A_0({\bf a'}))|{p}_{S}^{d}(B_0({\bf b})){\langle A({\bf a'})B({\bf b})\rangle}_{S}
+ {p}_{S}^{d}(B_0({\bf b'})){\langle A({\bf a'})B({\bf b'})\rangle}_{S}|\le 2. \label{chsh_gsass}
\end{eqnarray}
Eq. (\ref{chsh_gsass}) contains explicitly four detection probabilities and four conditional expectation values. The latter can be calculated by using the rules of QM because of Ax. 2 in Sect. \ref{modello}, and formally coincide with expectation values of QM. If one puts them into Eq. (\ref{chsh_gsass}) the inequality can be interpreted as a condition that must be fulfilled by the detection probabilities in the ESR model. It is then important to stress that this condition implies that the detection probability $p_{S}^{d}(F)$ in Eq. (\ref{formuladipartenza}) cannot be equal to $1$ for every state $S$ and property $F$, hence the physical predictions of the ESR model are necessarily different from those of QM (see (b) at the end of Sect. \ref{consistenza}). Of course, we have as yet no theory allowing us to calculate precise values for $p_{S}^{d}(F)$. Nevertheless, should one be able to perform measurements that are close to ideality, the detection probabilities could be determined experimentally and then inserted into Eq. (\ref{chsh_gsass}). Two possibilities occur.

(i) There exist states and observables such that the conditional expectation values violate Eq. (\ref{chsh_gsass}). In this case one must reject the ESR model (hence R and LOC), or the set of additional assumptions introduced to attain Eq. (\ref{corrfunc_ass}), or both.

(ii) For every choice of states and observables the conditional expectation values fit in with Eq. (\ref{chsh_gsass}). In this case the ESR model is supported by the experimental data.

The above alternatives show explicitly that the ESR model is, in principle, falsifiable, as we have stated at the end of Sect. \ref{consistenza}.\footnote{The implications of Eq. (\ref{chsh_gsass}) discussed above can be better understood by studying particular examples. In the case of two spin--$\frac{1}{2}$ quantum particles in the singlet spin state we have proven \cite{s07} that the  ESR model predicts, under suitable assumptions, the upper bound $\frac{1}{\sqrt[4]{2}}\approx 0.841$ for the probability that a spin--$\frac{1}{2}$ particle be detected when the spin along an arbitrary direction is measured on it.}

\section{A conciliatory result \label{unfairsampling}}
We have observed at the beginning of Sect. \ref{consistenza} that the ESR model aims to provide a set--theoretical picture of the physical world in which macroscopic properties can be considered objective. The basic elements of this picture are microscopic properties (primitive notion) and microscopic states (derived notion). We intend to broaden such picture in the present section by introducing theoretical probabilities of microscopic properties and microscopic observables. This broadening will be used to show that the standard BCHSH inequality, the modified BCHSH inequality and the standard quantum predictions do not conflict in the framework of the ESR model because they refer to different parts of the picture provided by the model. %Furthermore, we will also prove that the violation of the standard BCHSH inequalities predicted by the ESR model at a macroscopic level can be explained in terms of a (unconventional) kind of \emph{unfair sampling}. 

To begin with, let us recall from Sect. \ref{modello} that, whenever an ensemble $\Sigma$ of physical objects is prepared in a state $S$, the microscopic properties possessed by each object depend on the microscopic state $S^{i}$ of the object but not on the measurement context. It follows that, for every $f  \in \mathscr E$, one can introduce a theoretical probability ${p}_{S}(f)$ that a physical object $x$ in the state $S$ possess $f$. Furthermore, let us consider the macroscopic property $F= \varphi (f)$ corresponding to $f$. The probability ${p}_{S}(f)={p}_{S}(\varphi^{-1}(F))$ generally does not coincide with the probability ${p}_{S}^{t}(F)$ in Eq. (\ref{formuladipartenza}) because there may be physical objects that possess $f$ and yet are not detected, which implies that they do not display $F$ (hence, ${p}_{S}^{t}(F) \le {p}_{S}(f)$). As far as ${p}_{S}(f)$ and ${p}_{S}(F)$ are concerned, instead, two possibilities occur.

(i) The subensemble $\Sigma^{d}$ of all physical objects that are detected is a \emph{fair sample} of $\Sigma$, that is, the percentage of physical objects possessing $f$ in $\Sigma^{d}$ is identical to the percentage of physical objects possessing $f$ in $\Sigma$. Since all detected objects possessing $f$ turn out to display $F=\varphi(f)$ when a measurement is done, ${p}_{S}(f)$ and ${p}_{S}(F)$ coincide.

(ii) $\Sigma^{d}$ is not a fair sample of $\Sigma$. In this case ${p}_{S}(f)$ does not coincide with ${p}_{S}(F)$. 

Let us now introduce microscopic observables and their expectation values in the  ESR model, as follows.

Let $A_0$ be a discrete generalized observable and let us use the symbols introduced in Sect. \ref{disgeneralizzate}. Hence $A_0$ is characterized by the macroscopic properties $F_0= (A_0, \{a_0 \})$, $F_1= (A_0, \{a_1 \})$, $F_2= (A_0, \{a_2 \})$, \ldots The property $F_0$ has no microscopic counterpart, while $F_1$, $F_2$, \ldots correspond to the microscopic properties $f_1=\varphi^{-1}(F_1)$, $f_2=\varphi^{-1}(F_2)$, \ldots, respectively. Then, we define the microscopic observable $\mathbb A$ corresponding to $A_0$ by means of the family $\{f_n \}_{n \in {\mathbb N}}$. The possible values of ${\mathbb A}$ are the outcomes $a_1, a_2, \ldots$ and its expectation value $\langle {\mathbb A} \rangle_{S}$ in the state $S$ is given by
\begin{equation}
\langle {\mathbb A} \rangle_{S}=\sum_{n} a_n {p}_{S}(f_n),
\end{equation}
where ${p}_{S}(f_n)$ is the theoretical probability of the microscopic property $f_n$. 

We are thus ready to discuss what is going on at a microscopic level. Indeed, by using the above definition we can consider the (dichotomic) microscopic observables ${\mathbb A}({\bf a})$, ${\mathbb A}({\bf a'})$, ${\mathbb B}({\bf b})$ and ${\mathbb B}({\bf b'})$, each of which has possible values $-1$ and $+1$, corresponding to the (trichotomic) macroscopic observables $A_0({\bf a})$, $A_0({\bf a'})$, $B_0({\bf b})$ and $B_0({\bf b'})$ introduced in Sect. \ref{disgeneralizzate}, respectively. Since all microscopic properties are either possessed or not possessed by a given physical object, the usual procedures leading to Eq. (\ref{chsh_69}) can be applied. Hence we get the standard BCHSH inequality, with $P({\bf a}, {\bf b})$, $P({\bf a}, {\bf b'})$, $P({\bf a'}, {\bf b})$ and $P({\bf a'}, {\bf b'})$ reinterpreted  in terms of microscopic observables. 

Bearing in mind our results in Sect. \ref{modifiedBCHSH}, we can draw the conclusion that, under suitable assumptions on the observables that are taken into account, different inequalities hold for different parts of the picture provided by the ESR model. 

(a) The standard BCHSH inequality holds at a microscopic level (which is purely theoretical and cannot be experimentally checked).

(b) The modified BCHSH inequality holds at a macroscopic level whenever all physical objects that are prepared are considered (which can be experimentally checked, at least in principle, see Sect. \ref{modifiedBCHSH}).

(c) The quantum predictions deduced by using QM rules hold at a macroscopic level whenever only the physical objects that are detected are considered (which can be experimentally checked). In this case there are physical situations in which quantum inequalities hold which do not coincide with the standard BCHSH inequalities.%\footnote{For instance, the inequality
%\begin{displaymath}
%|{\bf a} \cdot {\bf b}- {\bf a} \cdot {\bf b'}|+|{\bf a'} \cdot {\bf b} + {\bf a'} \cdot {\bf b'}| \le 2 \sqrt{2}
%\end{displaymath}
%which holds in the specific case studied at the end of Sect. \ref{disgeneralizzate} \cite{se88a,hs91}.}

The above conclusion is ``conciliatory'' in the sense that it settles the conflict between the standard BCHSH inequality and quantum predictions, as anticipated at the beginning of this section. It is then interesting to observe that the ESR model allows us to explain the violation of the standard BCHSH inequality which occurs whenever quantum expectation values are substituted in this inequality in terms of a (unconventional) kind of unfair sampling. Indeed, let us suppose that $A_0$ is measured on each physical object in $\Sigma$. Then, several physical objects display the property $F_0$ (hence the expectation value $\langle A_0 \rangle_{S}$ of $A_0$ is given by Eq. (\ref{expectation_value})). Therefore the objects for which the outcomes $a_1, a_2, \ldots$ are obtained belong to the subset $\Sigma^{d} \subseteq \Sigma$. Furthermore, the probabilities ${p}_{S}(F_1)={p}_{S}(a_1)$, ${p}_{S}(F_2)={p}_{S}(a_2)$, \ldots must be interpreted as the large number limits of the frequencies of $a_1, a_2, \ldots$, respectively, in $\Sigma^{d}$. Let us consider the conditional expectation value $\langle A \rangle_{S}=\sum_{n} a_n {p}_{S}(F_n)$ introduced in Sect. \ref{disgeneralizzate} and compare it with $\langle {\mathbb A} \rangle_{S}$. It is apparent that $\langle A \rangle_{S}$ and $\langle {\mathbb A} \rangle_{S}$ must coincide if case (i) occurs, while they generally do not coincide if case (ii) occurs. Analogous remarks hold if we consider the conditional expectation value $\langle AB \rangle_S$ defined by Eq. (\ref{corrfunc_QM}). It follows that, if we substitute $P({\bf a}, {\bf b})$, $P({\bf a}, {\bf b'})$, $P({\bf a'}, {\bf b})$ and $P({\bf a'}, {\bf b'})$ in Eq. (\ref{chsh_69}) with the conditional expectation values ${\langle A({\bf a})B({\bf b})\rangle}_{S}$,  ${\langle A({\bf a})B({\bf b'})\rangle}_{S}$,  ${\langle A({\bf a'})B({\bf b})\rangle}_{S}$ and  ${\langle A({\bf a'})B({\bf b'})\rangle}_{S}$, respectively, the inequality must be fulfilled in case (i), while it can be violated in case (ii). Since the conditional expectation values coincide with standard quantum expectation values (see Sect. \ref{disgeneralizzate}), there are physical situations in which the Eq. (\ref{chsh_69}) is violated, hence we conclude that case (ii) occurs and $\Sigma^{d}$ is not a fair sample of $\Sigma$.\footnote{This explanation of the violation of the standard BCHSH inequality was already provided in \cite{gp04}, where however only macroscopic properties were considered and the distinction between a macroscopic property $F$ and its microscopic counterpart $f= \varphi^{-1}(F)$ was not explicitly introduced. Our argument was therefore somewhat ambiguous, and our present treatment also aims to amend this shortcoming. We add that unfair sampling obviously represents a necessary but not sufficient condition for the violation of Eq. (\ref{chsh_69}), so that further quantitative conditions on it must be imposed if this violation has to occur. For the sake of brevity, we do not discuss this topic here.}

We have thus attained our goal. To close up, we would like to add a final remark. It is well known that the tests of Bell's inequalities actually check derived inequalities, obtained by adding additional assumptions to \emph{local realism}. Therefore many scholars uphold that the experimental data that disprove these inequalities could actually show that the additional assumptions are false, not that local realism is untenable (see, \emph{e.g.}, \cite{f82,f89,s04,s05,dcg96,s00,sf02,gg99}). In addition, some authors point out that the proof of Bell's inequality requires a \emph{hidden Bell's postulate} (HBP) besides local realism, \emph{i.e.}, the assumption that ``an experiment involving several incompatible measurements can be written on a single probability space, independently of the measurement context'' \cite{a09}. HBP implies a \emph{fair sampling assumption} on the measuring apparatuses, which has been recently questioned by reconsidering some available experimental data \cite{ak07}; moreover, a wave model has been devised in which unfair sampling occurs \cite{a09}. These results provide further support to the V\"{a}xj\"{o} interpretation of QM, which rejects HBP and is contextual, statistical and realistic \cite{k07}. It is then interesting to observe that the foregoing criticism to the fair sampling assumption rests on investigations into real measuring processes, and unfair sampling is ascribed to features of the measuring apparatuses (\emph{e.g.}, existence of thresholds) rather than to intrinsic properties of the physical objects that are considered, as in the ESR model (where therefore no hidden variable associated with measuring apparatuses occurs, see Sect. \ref{modello}). This makes the V\"{a}xj\"{o} interpretation of QM basically different from the ESR model, which provides a noncontextual interpretation and generalization of QM and yet predicts that the standard BCHSH inequality is violated in both cases (b) and (c) that can occur at a macroscopic level, though it holds at a microscopic, purely theoretical level, as we have seen above.


\begin{thebibliography}{99}

\bibitem{blm91} Busch, P., Lahti, P.J., Mittelstaedt, P.: The Quantum Theory of Measurement. Springer, Berlin (1991)

\bibitem{b66} Bell, J.S.: On the problem of hidden variables in quantum mechanics. Rev. Mod. Phys. \textbf{38}, 447--452 (1966)

\bibitem{ks67} Kochen, S., Specker, E.P.: The problem of hidden variables in quantum mechanics. J. Math. Mech. \textbf{17}, 59--87 (1967)

\bibitem{b64} Bell, J.S.: On the Einstein-Podolsky-Rosen paradox. Physics \textbf{1}, 195--200 (1964)

\bibitem{bs96} Busch, P., Shimony, A.: Insolubility of the quantum measurement problem for unsharp observables. Stud. His. Phil. Mod. Phys. \textbf{27B}, 397--404 (1996)

\bibitem{m93} Mermin, N.D.: Hidden variables and the two theorems of John Bell. Rev. Mod. Phys. \textbf{65}, 803--815 (1993)

\bibitem{gs04} Garola, C., Sozzo, S.: A semantic approach to the completeness problem in quantum mechanics. Found. Phys. \textbf{34}, 1249--1266 (2004)

\bibitem{ga00} Garola, C.: Objectivity versus nonobjectivity in quantum mechanics. Found. Phys. \textbf{30}, 1539--1565 (2000)

\bibitem{gs96a} Garola, C., Solombrino, L.: The theoretical apparatus of semantic realism: a new language for classical and quantum physics. Found. Phys. \textbf{26}, 1121--1164 (1996)

\bibitem{gs96b} Garola, C., Solombrino, L.: Semantic realism versus EPR-like paradoxes: the Furry, Bohm-Aharonov, and Bell paradoxes. Found. Phys. \textbf{26}, 1329--1356 (1996)

\bibitem{ga02b} Garola, C.: Essay review: Waves, Information, and Foundations of Physics. Stud. Hist. Phil. Mod. Phys. \textbf{33}, 101--116 (2002)

\bibitem{ga02} Garola, C.: A simple model for an objective interpretation of quantum mechanics. Found. Phys. \textbf{32}, 1597--1615 (2002)

\bibitem{gp04} Garola, C., Pykacz, J.: Locality and measurements within the SR model for an objective interpretation of quantum mechanics. Found. Phys. \textbf{34}, 449--475 (2004)

\bibitem{ga03} Garola, C.: Embedding quantum mechanics into an objective framework. Found. Phys. Lett. \textbf{16}, 605--612 (2003)

\bibitem{ga05} Garola, C.: MGP versus Kochen-Specker condition in hidden variables theories. Int. J. Theor. Phys. \textbf{44}, 807--814 (2005)

\bibitem{gps06} Garola, C., Pykacz, J., Sozzo, S.: Quantum machine and semantic realism approach: a unified model. Found. Phys. \textbf{36}, 862--882 (2006)

\bibitem{chsh69} Clauser, J.F., Horne, M.A., Shimony, A., Holt, R.A.: Proposed experiment to test local hidden-variable theories. Phys. Rev. Lett. \textbf{23}, 880--884 (1969)

%\bibitem{gs07b} Garola, C., Sozzo, S.: Reinterpreting quantum probabilities in a realistic and local framework: the modified BCHSH inequalities. ArXiv:quant-ph/0703260v5 (2007)

\bibitem{g07} Garola, C.: The ESR model: reinterpreting quantum probabilities within a realistic and local framework. In Adenier, G., et al. (eds) Quantum Theory: Reconsideration of Foundations 4, pp. 247--252. American Institute of Physics, Melville, New York (2007)

\bibitem{bc81} Beltrametti, E.G., Cassinelli, G.: The Logic of Quantum Mechanics. Addison--Wesley, Reading, MA (1981)

\bibitem{l83} Ludwig, G.: Foundations of Quantum Mechanics I. Springer, Berlin (1983)

\bibitem{aa04} Aerts, D., Aerts, S.: Towards a general operational and realistic framework for quantum mechanics and relativity theory. In: Elitzur, A.C., Dolev, S., Kolenda, N. (eds.) Quo Vadis Quantum Mechanics? Possible Developments in Quantum Theory in the 21st Century. Springer, Berlin (2004)

\bibitem{k07} Khrennikov, A.Y.: Contextual Approach to Quantum Formalism. Springer, Berlin (2009)

\bibitem{sg08} Sozzo, S., Garola, C.: A Hilbert space representation of generalized observables and measurement processes in the ESR model. ArXiv:0811.0531v2 [quant-ph]. Submitted to Int. J. Theor. Phys.

\bibitem{gs09a} Garola, C., Sozzo, S.: The ESR model: a proposal for a noncontextual and local Hilbert space extension of QM. Europhys. Lett. \textbf{86}, 20009p1--20009p6 (2009)

\bibitem{epr35} Einstein, A., Podolsky, B., Rosen, N.: Can quantum-mechanical description of physical reality be considered complete?. Phys. Rev. \textbf{47}, 777--780 (1935)

\bibitem{n07} Norsen, T.: Against `Realism'. Found. Phys. \textbf{37}, 311--340 (2007)

\bibitem{s07} Sozzo, S.: Modified BCHSH inequalities within the ESR model. In: Adenier, G., et al. (eds) Quantum Theory: Reconsideration of Foundations 4, pp. 334--338. American Institute of Physics, Melville, New York (2007)

\bibitem{f82} Fine, A.: Hidden variables, joint probability and the Bell inequalities. Phys. Rev. Lett. \textbf{48}, 291--295 (1982)

\bibitem{f89} Fine, A.: Correlations and efficiency: testing the Bell inequalities. Found. Phys. \textbf{19}, 453--478 (1989)

\bibitem{s04} Santos, E.: The failure to perform a loophole-free test of Bell's inequality supports local realism. Found. Phys. \textbf{34}, 1643--1673 (2004)

\bibitem{s05} Santos, E.: Bell's theorem and the experiments: increasing empirical support for local realism?. Stud. Hist. Phil. Mod. Phys. \textbf{36}, 544--565 (2005)

\bibitem{dcg96} De Caro, L., Garuccio, A.: Bell's inequality, trichotomic observables, and supplementary assumptions. Phys. Rev. A \textbf{54}, 174--181 (1996) 

\bibitem{s00} Szabo, L.E.: On Fine's resolution of the EPR-Bell problem. Found. Phys. \textbf{30}, 1891--1909 (2000)

\bibitem{sf02} Szabo, L.E., Fine, A.: A local hidden variable theory for the GHZ experiment. Phys. Lett. A \textbf{295}, 229--240 (2002)

\bibitem{gg99} Gisin, N., Gisin, B.: A local hidden variable model of quantum correlation exploiting the detection loophole. Phys. Lett. A \textbf{260}, 323--327 (1999)

\bibitem{a09} Adenier, G.: Violation of Bell inequalities as a violation of fair sampling in threshold detectors. In: Accardi, L., et al. (eds) Foundations of Probability and Physics-5, pp. 8--17. American Institute of Physics, Melville, New York (2009)

\bibitem{ak07} Adenier, G., Khrennikov, A.Y.: Is the fair sampling assumption supported by the EPR experiments?. J. Phys. B \textbf{42}, 131--141 (2007) 

\end{thebibliography}
\end{document}